\documentclass[journal=jacsat,manuscript=article]{achemso}
\setkeys{acs}{articletitle = true}

\usepackage{array}
\usepackage{booktabs}
\usepackage[font=small,labelfont=bf]{caption}
\usepackage[version=3]{mhchem} 
\usepackage{hyperref}
\usepackage{graphicx}
 \graphicspath{ {./} }
\usepackage{epstopdf}
\usepackage[compatibility=false]{caption}
\DeclareCaptionFont{quackfont}{\fontfamily{ptm}\fontsize{9pt}{9pt}\selectfont}
\usepackage[font=quackfont]{subcaption}
\usepackage{cleveref}
\author{Aditya N. Singh}
\author{Arun Yethiraj}
\email{yethiraj@wisc.edu}
\affiliation[UW-Madison]
{Theoretical Chemistry Institute and Department of Chemistry, 1101 University Avenue, University of Wisconsin-Madison, Madison, Wisconsin 53703}

\title[An \textsf{achemso} demo]
  {Liquid-liquid Phase Separation as the Second Step of Complex Coacervation}

\abbreviations{pY,pR}

\begin{document}
\begin{tocentry}
\includegraphics[width=\linewidth]{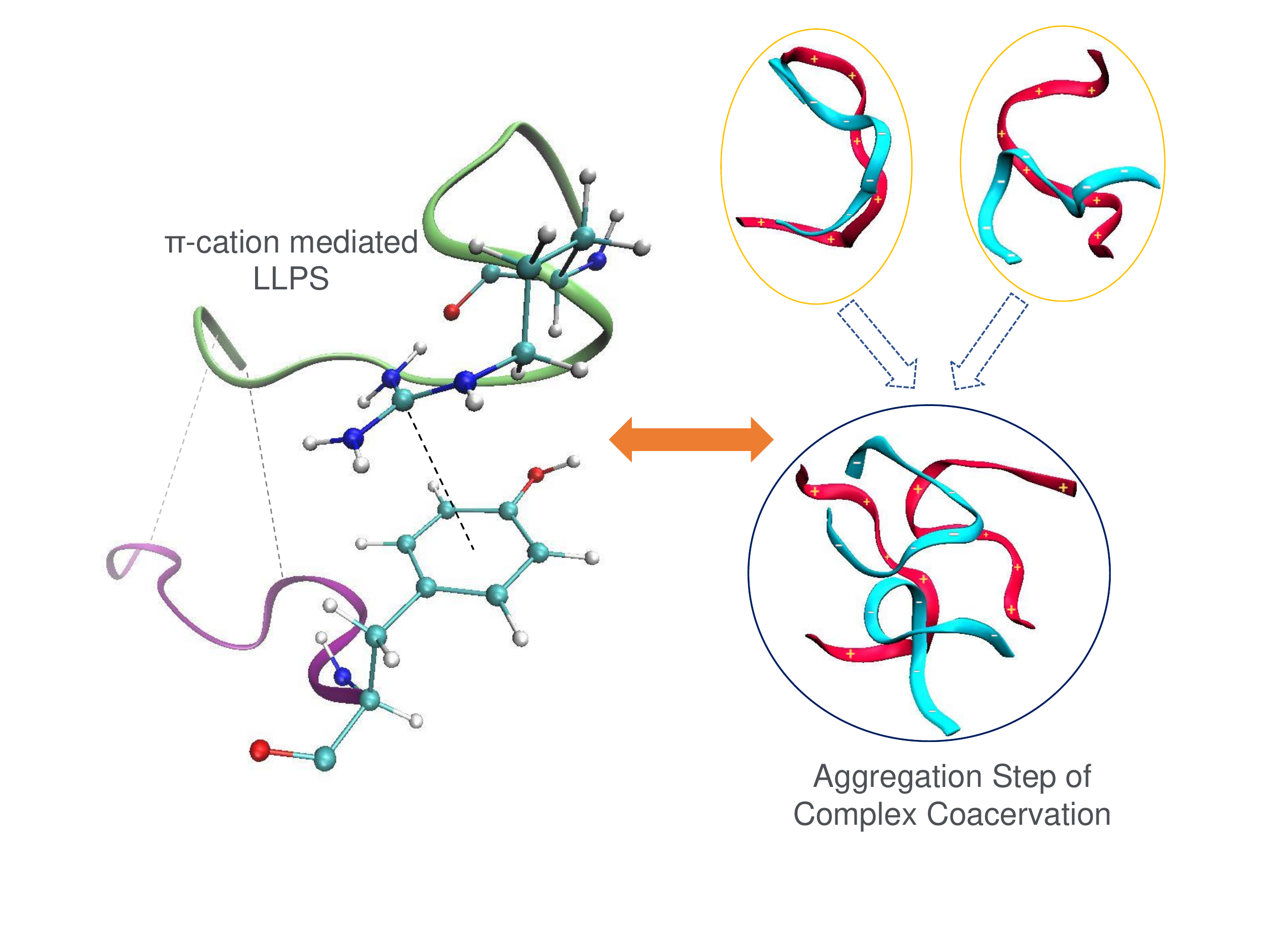}

\end{tocentry}
\begin{abstract}
Liquid liquid phase separation (LLPS) mediated by $\pi$-cation bonds between tyrosine and arginine residues are of biological importance. To understand the interactions between proteins in the condensed phase in close analogy to complex coacervation, we run multiple umbrella calculations between oligomers containing tyrosine (pY) and arginine (pR). We find pR-pY complexation to be energetically driven. Metadynamics simulations reveal that this energy of complexation comes primarily from $\pi$-cation bonds. On running free energy calculation for the second binding step of complex coacervation, we find striking similarities between this process and $\pi$-mediated LLPS. These calculations lead us to believe that contrary to the common notion, complex coacervation as whole, which involves an entropic complexation followed by an energetic aggregation is not invoked by proteins containing arginine and tyrosine residues. Rather, the latter step in itself, in which neutral polyion pairs aggregate together is the correct mechanism for $\pi$-cation mediated LLPS. 
\end{abstract} 

\section{Introduction}

As early as the 1930s, Alexander Oparin proposed that the first step in the origin of life was the phase separation of macromolecules into liquid coacervates (or protein-dense liquid droplets)\cite{oparin1957origin}.   The last decade has seen an explosion of research on liquid-liquid phase separation (LLPS), the phenomenon in which proteins or nucleic acids condense into a dense phase (resembling liquid droplets) and a dilute phase\cite{alberti2019considerations}. Multiple studies have established that this phenomenon underlies the ubiquitous formation of membraneless compartments in tcells.\cite{larson2017liquid,schmidt2016transport,su2016phase,banani2017biomolecular,shin2017liquid}   LLPS has also been linked to a range of diseases such as amyotoriphic lateral sclerosis (ALS)\cite{alberti2019considerations}, Alzheimer's and Parkinson's disease\cite{alberti2019liquid} and pancreatic cancer\cite{xu2019targeting}. The prevalence of LLPS in biological systems as well as its application in drug delivery\cite{doi:10.1517/17425247.2014.941355} and discovery\cite{indulkar2016exploiting}, protein purification\cite{Wang1996356} and artificial cell development\cite{crowe2018liquid} is a strong motivation to understand the thermodynamics underlying this process.

There is a strong parallel between LLPS in cells and phase transitiopns in polymer solutions\cite{perry2019phase}. 
It has been proposed that in some biological systems, such as the Fused in Sarcoma (FUS) protein and the Nephrin Intracellular Domain (NICD), the LLPS is driven by complex coacervation, a popular phenomenon observed in polyelectrolyte systems.\cite{perry2019phase,qamar2018fus,pak2016sequence,wang2018molecular} 
Complex coacervation is a process in which oppositely-charged polyelectrolytes in aqueous solution undergo a liquid-liquid phase separation to form a polymer-dense (or coacervate) and a polymer-dilute  phase.\cite{decher1997fuzzy,van2011polyelectrolyte}.  The properties of coacervates can be tuned by the charge density\cite{doi:10.1002/jcp.1030490404} and chirality\cite{C4SM02336F} of the polyelectrolytes, pH\cite{doi:10.1021/bm300835r},  ionic strength\cite{PhysRevLett.105.208301}, temperature\cite{doi:10.1021/ma902144k} and the concentration of salt\cite{doi:10.1002/jcp.1030490404}, allowing for the possibility of understanding LLPS from changes in the physical environment.

In this work we study the association between poly-tyrosine and poly-arginine, which are residues prominent in LLPS and compare to the association between poly-glutamate and poly-lysine, which is an example complex coacervation.  

While the analogy between complex coacervation and LLPS is interesting, we note that in the case of complex coacervation the two polyion species have opposite charge (each neutralized by their respective counterions), which the LLPS in FUS and NICD is mediated by the $\pi$-cation bonds between arginine and (electrically neutral) tyrosine residues\cite{wang2018molecular,pak2016sequence,qamar2018fus,kang2019unified}. 
Furthermore, the thermodynamics of complex coaccervation in itself is a controversial topic\cite{sing2017development,van2011polyelectrolyte,rathee2018role}. One hypothesis is that complex coacervation is entropically favorable and driven by the gain in the translational entropy of small ions when they are released from being bound or 'condensed' to the polyelectrolytes\cite{DEKRUIF2004340,gummel2007,PRIFTIS201339,doi:10.1021/la100705d,doi:10.1063/1.2178803,rathee2018role,doi:10.1021/bm201113y}. Other arguments emphasize the importance of water pertubation \cite{doi:10.1021/jacs.5b11878}, dipole-dipole interactions \cite{adhikari2018polyelectrolyte} and increased entropy of having multiple binding partners\cite{lytle2018tuning}.

A two-step formulation of complex coacervation has been proposed\cite{priftis2012thermodynamic}  that bridges the apparent gap between LLPS and complex coacervation.  The first step is the complexation of oppositely charged polyions, to make a neutral complex; a process that is entropically favorable due to counterion release.  The second step is the phase separation of solutions of neutral complexes into polymer rich and polymer poor phases,; a process that is energetically favorable.  The first step is supported by multiple studies.\cite{singh2020driving, doi:10.1063/1.2178803,rathee2018role} The second step results in an upper critical solution temperature and has been treated using theories for phase transitions in neutral polymer solutions \cite{pappu-nature-physics}.

In this work, we use MD simulations and umbrella sampling to investigate the thermodynamics of LLPS in oligomers of poly-tyrosine (pY) and poly-arginine (pR).  They are compared to the umbrella sampling calculation between two neutral complexes (made up of paired poly-lysine and poly-glutamate) which provides a way to capture the thermodynamics of the second binding step of coacervation. Finally, metadynamics simulations are performed to discuss the nature of the $\pi$-cation interactions between pR and pY.

\section{Computational Methods}

\subsection{Umbrella Sampling and MD Simulation}

For the atomistic simulations, the AMBER ff14SB forcefield\cite{maier2015ff14sb} is used in this work along with TIP3P water.\cite{jorgensen1983comparison}. The ion parameters by Joung and Cheatham \cite{joung2008determination} are used to prevent crystallization in systems with high concentration of salt. The two oligomers used in this system are poly-tyrosine with 11 residues (pY) and poly-arginine with 6 residues of charged arginine and 5 residues of neutral arginine (pR). The scaled charges for neutral arginine are obtained from the B3LYP/cc-pVTZ charge scaling performed by Weis et al.\cite{weis2006ligand}. Four umbrella sampling calculations with 2 oligomers are performed, for the pairs pR-pR, pY-pY, pR-pY and pR-pY, all with 0.1 M excess salt.

Simulations are performed using the GROMACS 5.1.4\cite{ABRAHAM201519} package. The Lennard-Jones cutoff is set to 1 nm for the AMBER forcefield and 1.4 nm for the Martini forcefield.  The Particle Mesh Ewald\cite{darden1993particle} method is used to calculate the electrostatic interactions with the following configuration: for the AMBER forcefield, the real cutoff spacing is 1 nm and the fast Fourier transform grid spacing is 0.16 nm; for the Martini forcefield the real cutoff spacing is 1.4 nm and the fast Fourier transform grid spacing is 0.20 nm. 
The Berendsen barostat\cite{doi:10.1063/1.448118} is used to keep the pressure constant, and  the velocity rescaling algorithm \cite{bussi2007canonical} is used to keep the temperature constant. 

Initial configurations are created by inserting molecules randomly into in a cubic box of size 10x10x10 nm$^3$ with periodic boundary conditions in all directions.  The energy is minimized using a steepest decent algorithm, and the system is then equilibrated in the NPT ensemble at a pressure of 1 bar. The final configuration obtained from NPT equilibration is used for the pulling simulation in the NVT ensemble. 
The two polypeptides are pulled apart along the x-direction to generate multiple windows for the umbrella sampling simulations. 30 windows are used for a distance of separation between the central residue of the oligomers ($\xi$) from 0.4 to 3.8 nm. 

For the umbrella sampling production runs, a harmonic force constant of 1000 kJ mol${^{-1}}$ nm$^2$ is applied to constrain the distance of separation between the two polypeptides.   All production runs are done in a NVT ensemble.
Finally, the weighted histogram analysis method\cite{doi:10.1002/jcc.540130812}(WHAM) is employed to obtain the potential of mean force curves from the histograms. The last 75$\%$ of the production runs are used for WHAM analysis. The standard deviation for the PMF curves are computed by using a bootstrapping method in which complete histograms are considered as independent data points. To ensure that the system is equilibrated, the PMF obtained from the first 25$\%$ of the production run is compared to that obtained from the last 75$\%$ of the simulation run. The two potential of mean curves were within less than half a standard deviation of each other.

Using the method thus described, PMF curves are obtained for two temperatures, 280K and 320K. Assuming that the energy and entropy of association is constant between the temperature of 280K and 320K, the PMF curves obtained from umbrella sampling are decomposed into energetic($\Delta U(\xi$)) and entropic($\Delta S(\xi$)) contributions at a given distance of separation using the equations:
    \begin{equation}
        \Delta S(\xi) = - \frac{\Delta A(\xi, 320K) - \Delta A(\xi, 280K)}{(320K - 280K)}
       \label{eq:1}
    \end{equation}
\begin{equation}
    \Delta U(\xi) = \Delta A(\xi) + T\Delta S(\xi)
    \label{eq:2}
\end{equation}
Here $\Delta$A is the Hemholtz free energy which is numerically equal to the value of the shifted PMF curve. The standard deviation for $\Delta A$ is calculated by using bootstrap analysis. The error bars shown in the plots correspond to one standard deviation of the quantity of interest.

\begin{figure}[H]
\includegraphics[width=0.8\linewidth]{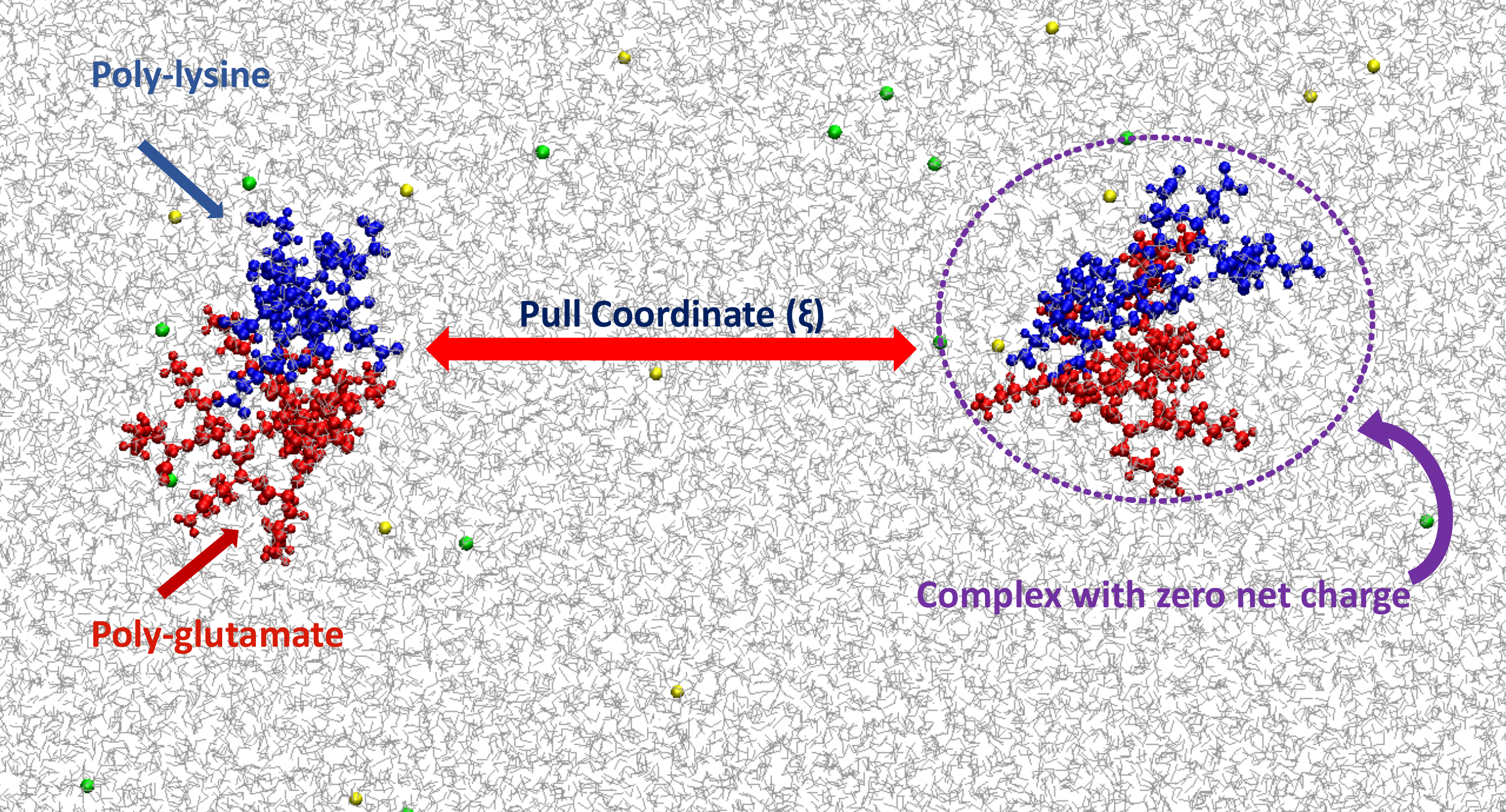}
\caption{Visualization of the system with four oligomers. The two neutral complexes each contain a poly-lysine and poly-glutamate oligomer, and the PMF is calculated between the center of mass of the two complexes.}
  \label{fig:4_chains_visual}
\end{figure}

One final umbrella sampling calculation is performed in a system that has 4 oligomers present - 2 oligomers of poly-lysine (10 residues) and 2 oligomers of poly-glutamate (10 residues). In this system two neutral complexes are formed, each of them between 1 oligomer of poly-lysine and 1 oligomer of poly-glutamate. The coordinate along which the PMF is calculated is the distance between center of mass between these two neutrally charged complexes (as visualized in figure \cref{fig:4_chains_visual}. The motivation behind this calculation is to develop the simplest system that can capture the second binding step of complex coacervation mentioned before. 
\subsection{Well Tempered Metadynamics} \label{ssec:metad}

\begin{figure}[H]
\includegraphics[width=0.8\linewidth]{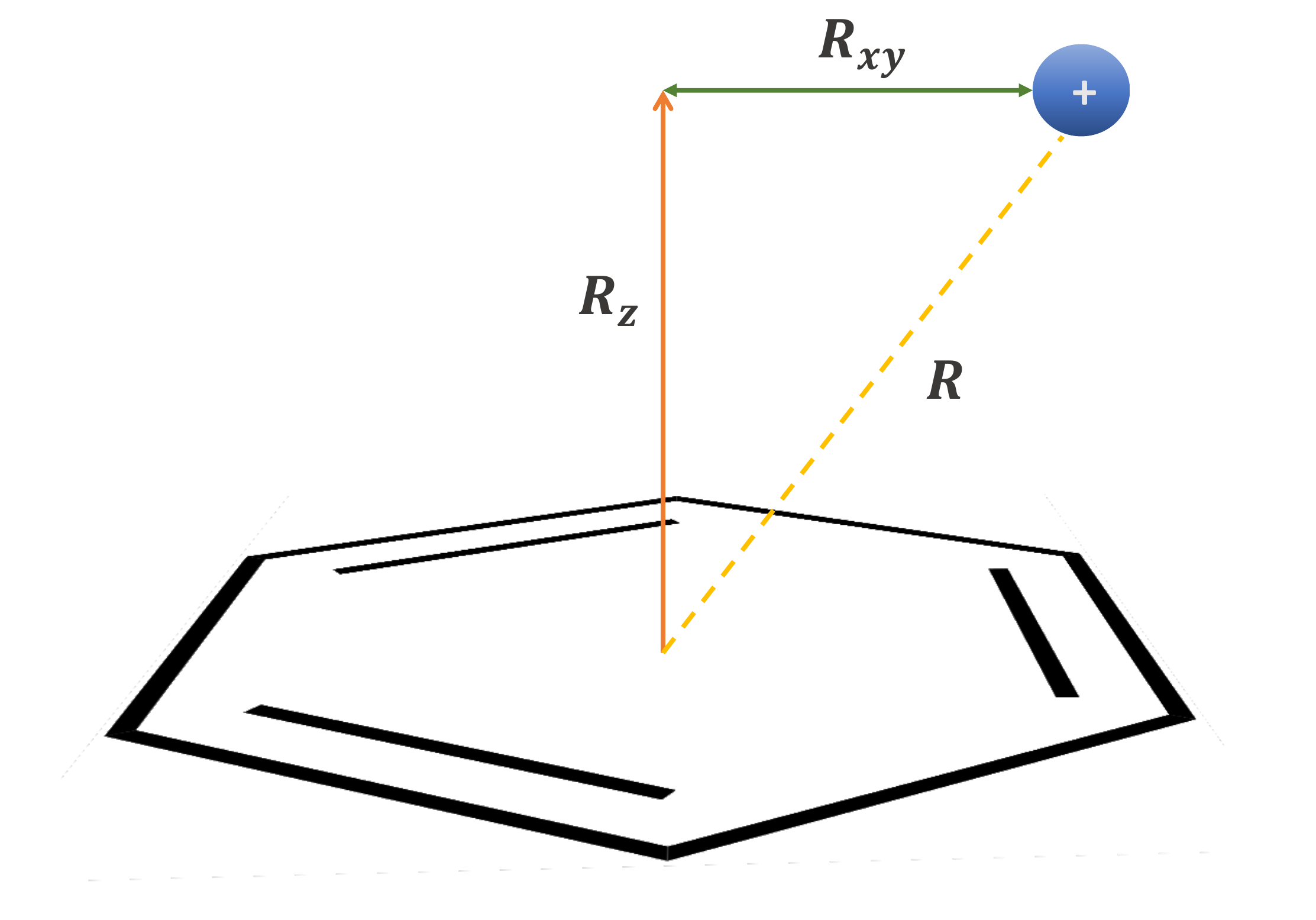}
\caption{The two collective variables chosen to quantify the $\pi$-cation bonding between pR and pY. $R$ is the distance between the arginine cation and the center of mass of the aromatic tyrosine ring. $R_z$ is the component of $R$ along the line perpendicular to the plane of the ring and $R_{xy}$ is the distance along either vertical or the horizontal axis.}
  \label{fig:picat_visual}
\end{figure}

The key idea of this calculation is to develop a quantitative measure of a $\pi$-cation bond. Recent developments by Kumar et al\cite{kumar2018cation} propose such a quantitative paradigm by performing gas phase DLPNO-CCSD(T)/aug-cc-pVTZ potential energy surface (PES) calculations and data-mining empirical PDB structures that contain $\pi$-cation bonds between residues of arginine and tyrosine. They do so by calculating the distance ($R$) between the carbocation on the arginine and the center of mass of the aromatic ring of the tyrosine and then splitting it into its components along the line perpendicular to the plane of the ring $R_z$ and the horizontal or vertical component ($R_{xy}$). These components are visualized in Figure \ref{fig:picat_visual}. Their analysis reveals that majority of empirical $\pi$-cation bonds between tyrosine and arginine are formed at $R_z < 0.6 $nm and $R_{xy} < 0.3 $nm.

The free energy of $\pi$-cation interactions between tyrosine and arginine in solution is obtained using 
2D Well-Tempered Metadynamics \cite{barducci2008well} with PLUMED v2.5\cite{tribello2014plumed}.  The system comprises of two oligomers - one with a single residue of arginine (R) and the other with a single residue of tyrosine (Y). Both oligomers are capped by an acetyl (ACE) in the N-terminal and N-methyl (NME) in the C-terminal to mask any strong interactions between terminals. The forcefield parameters used in the MD simulations are also used for metadynamics. 
The two collective variables used in this system are $R_z$ and $R_{xy}$ as used in discussed before, however the $R_{xy}$ is defined as $\sqrt{R_x^2 + R_y^2}= \sqrt{R^2 - R_z^2}$.   The simulation is run for 40ns and gaussians of height 2.0 Joule mol$^{-1}$ and $\sigma = 0.005$ nm are added at a stride of 1 ps. A bias factor of 5 is used for performing well-tempered metadynamics.

\section{Results and Discussion}

\subsection{Potential of mean-force}

\begin{figure}[H]
  \centering
  \begin{subfigure}[b]{\linewidth}
   \centering
    \includegraphics[width=0.475\linewidth]{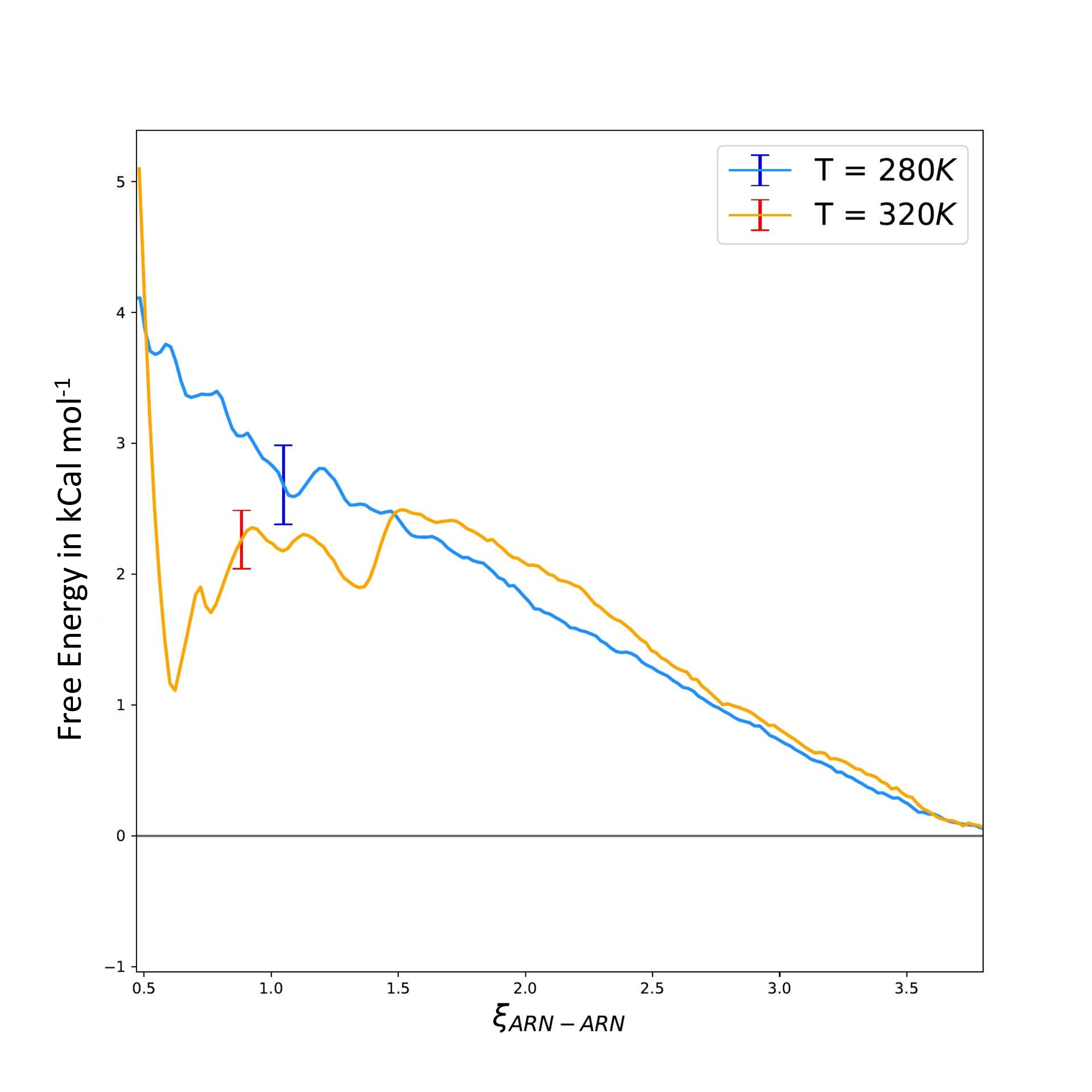}
    \hfill
    \includegraphics[width=0.475\linewidth]{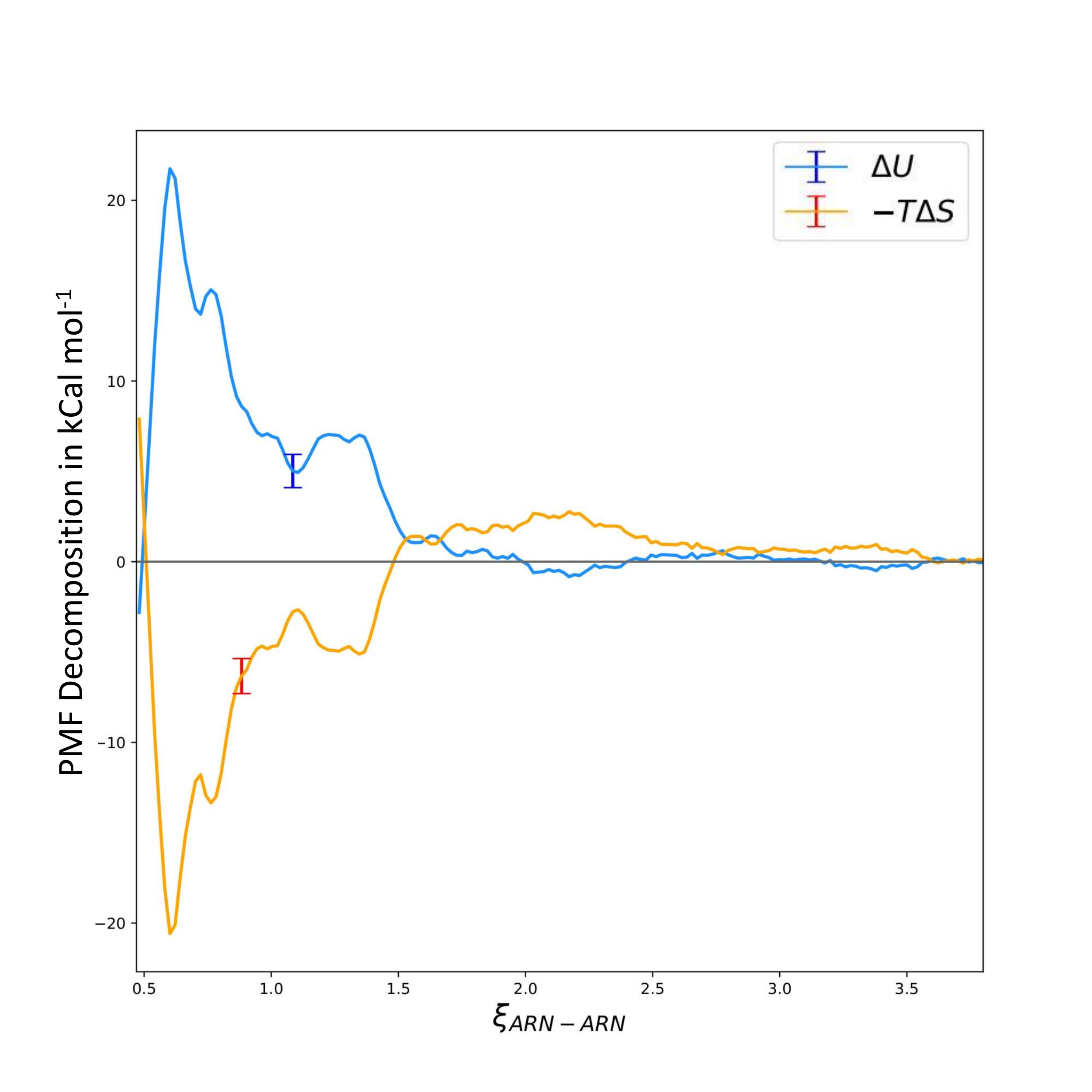}
    \caption{PMF (left) and its decomposition into its energetic and entropic terms (right) for the complexation between two oligomers of poly-arginine (\textbf{pR-pR}).}
  \end{subfigure}
   \begin{subfigure}[b]{\linewidth}
   \centering
    \includegraphics[width=0.475\linewidth]{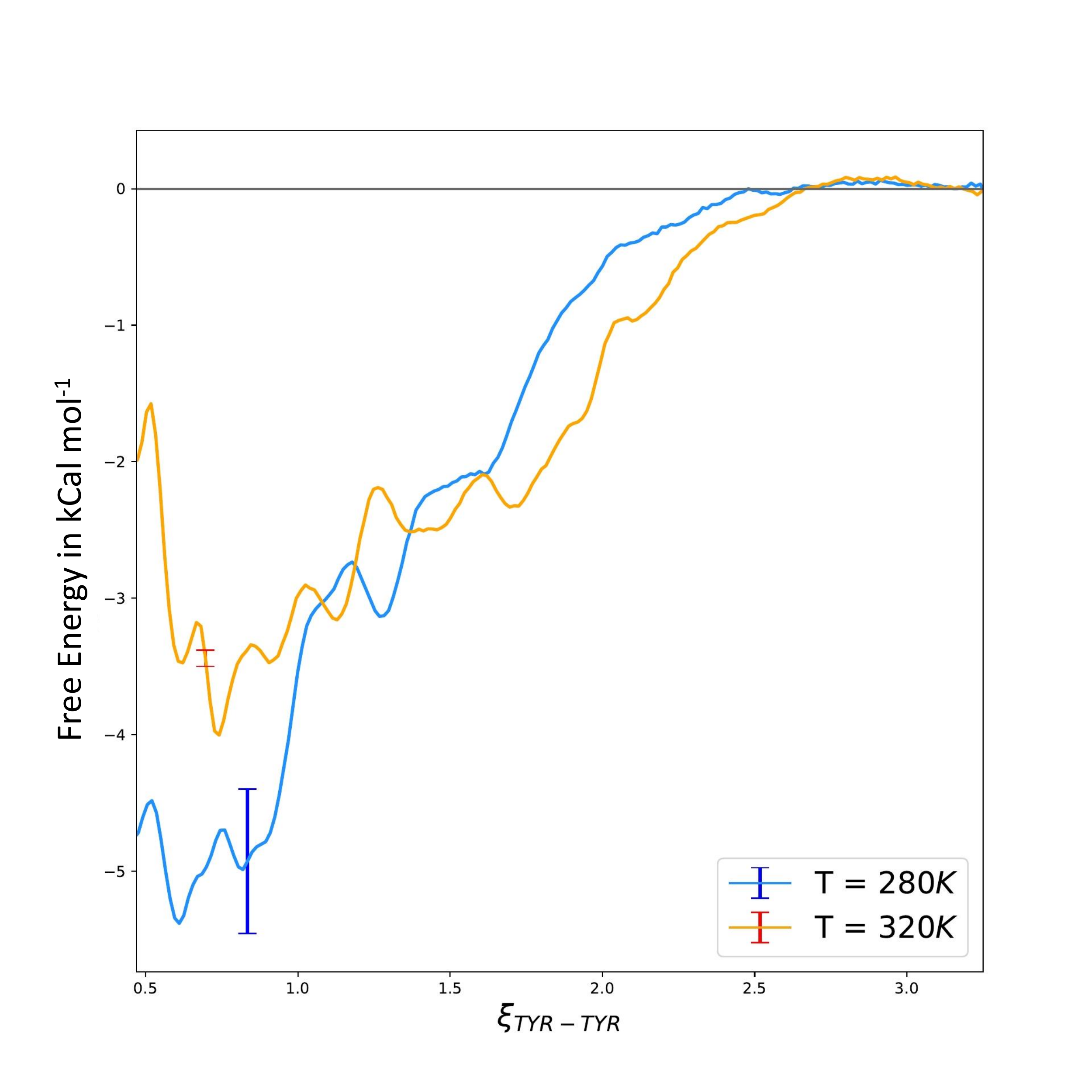}
    \hfill
    \includegraphics[width=0.475\linewidth]{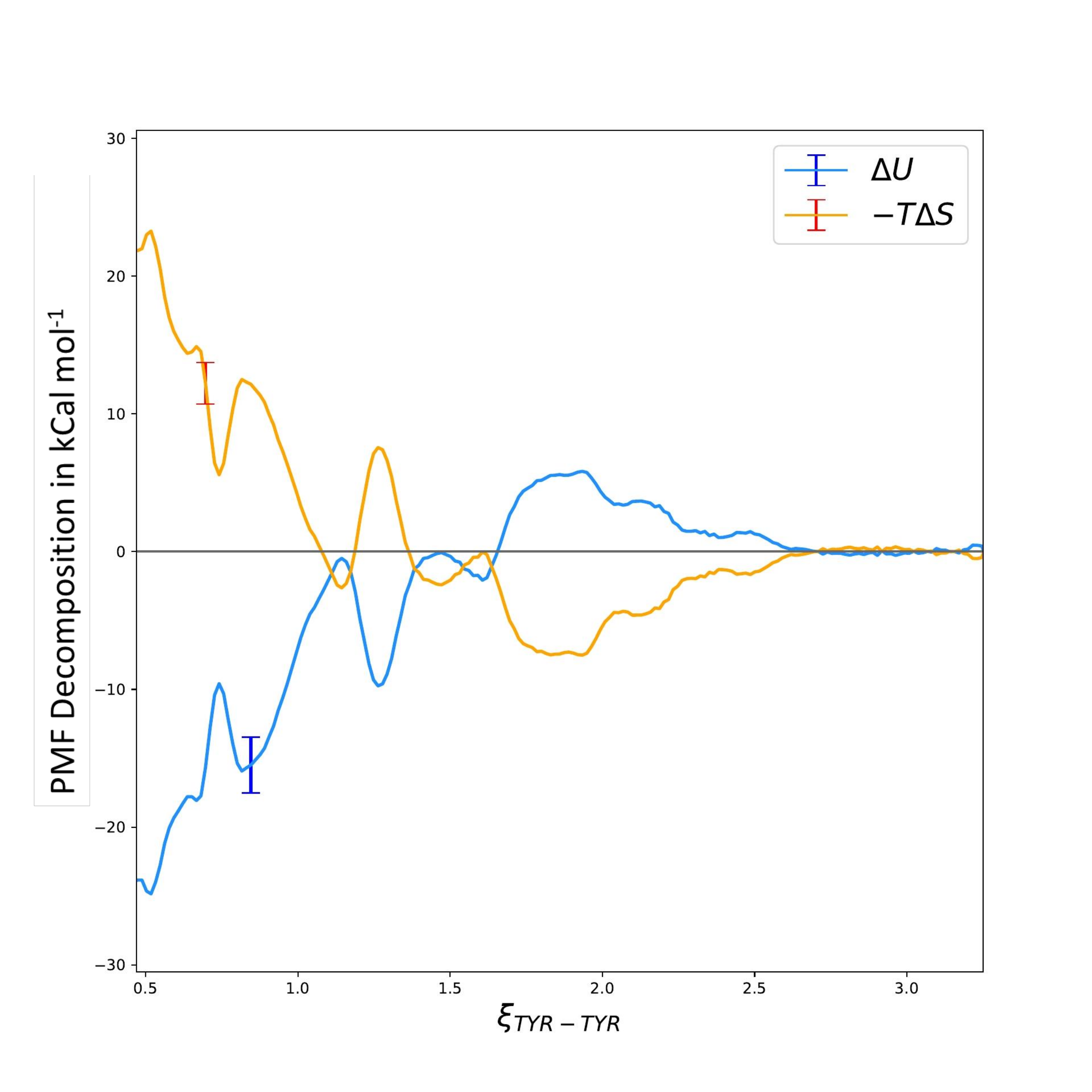}
    \caption{PMF (left) and its decomposition into its energetic and entropic terms (right) for the complexation between two oligomers of poly-tyrosine (\textbf{pY-pY}).}
    
  \end{subfigure}
  \caption{}
  \label{fig:like}
\end{figure}

The association between two pR oligomers is unfavorable, but that between two pY oligomers is favorable.  The potential of mean-force and its decomposition into energy and entropy for pR-pR is shown in Fig \ref{fig:like} (a). The free energy for this system is positive throughout the reaction coordinate ($\xi_{ARN-ARN}$), indicating that pR-pR association is unfavorable. This is expected since the net charge on both the polypeptides is +6. Hence this unfavorable interaction between pR-pR can be attributed to the electrostatic repulsion between them. One relevant observation is that a potential well is present in the pR-pR system at T=320 K and $\epsilon_{ARN-ARN}$ = 0.6 nm , but absent for T = 280 K.  Note also that over most of the range of the reaction co-ordinate the value of the PMF decreases as the temperature is increased, which implies that the entropy change is positive.

A decomposition of the pR-pR PMF into energy  and entropy shows that the complexation is entropically favorable.  In the case of the complexation of polymers of opposite change, the polyions neutralize each other and this releases the counterions, thus resulting in an increase in translational entropy.  In the case of pR-pR oligomers there is no counterion release, since both oligomers have the same charge.  In fact, the pair correlation function between oligomer and counterions is essentially unchanged as the reaction co-ordinate is changed.  There are changes, however, in the water structure and we speculate that the entropic component is due to water correlations, as is seen in other systems\cite{mondal-pmf}. 
The energetic component, which arises from the electrostatic repulsion is highly repulsive, overcoming the entropy of complexation and making association as a whole unfavorable. 

The association of pY oligomers, on the other hand, is favorable, and has an energetic driving force
(see Fig \ref{fig:like} (b)). The PMF for this system is both qualitatively and quantitatively different from the pR-pR system. The free energy of association at both temperatures is negative at the potential well indicating that pY-pY complexation is favorable. The free energy at the lower temperature is more favorable, indicating that complexation is exothermic or energetically driven. This is also clearly visible in the free energy decomposition plot. The favorable energetic contribution can be possibly attributed to the formation of stable hydrogen bonds between the tyrosine residues of the two oligomers, an interaction that has been found to drive FUS demixing.\cite{qamar2018fus} These bonds become unfavorable at high temperatures, due to which the magnitude of the free energy decreases with temperature.

\begin{figure}[H]
  \centering
   \centering
    \includegraphics[width=0.475\linewidth]{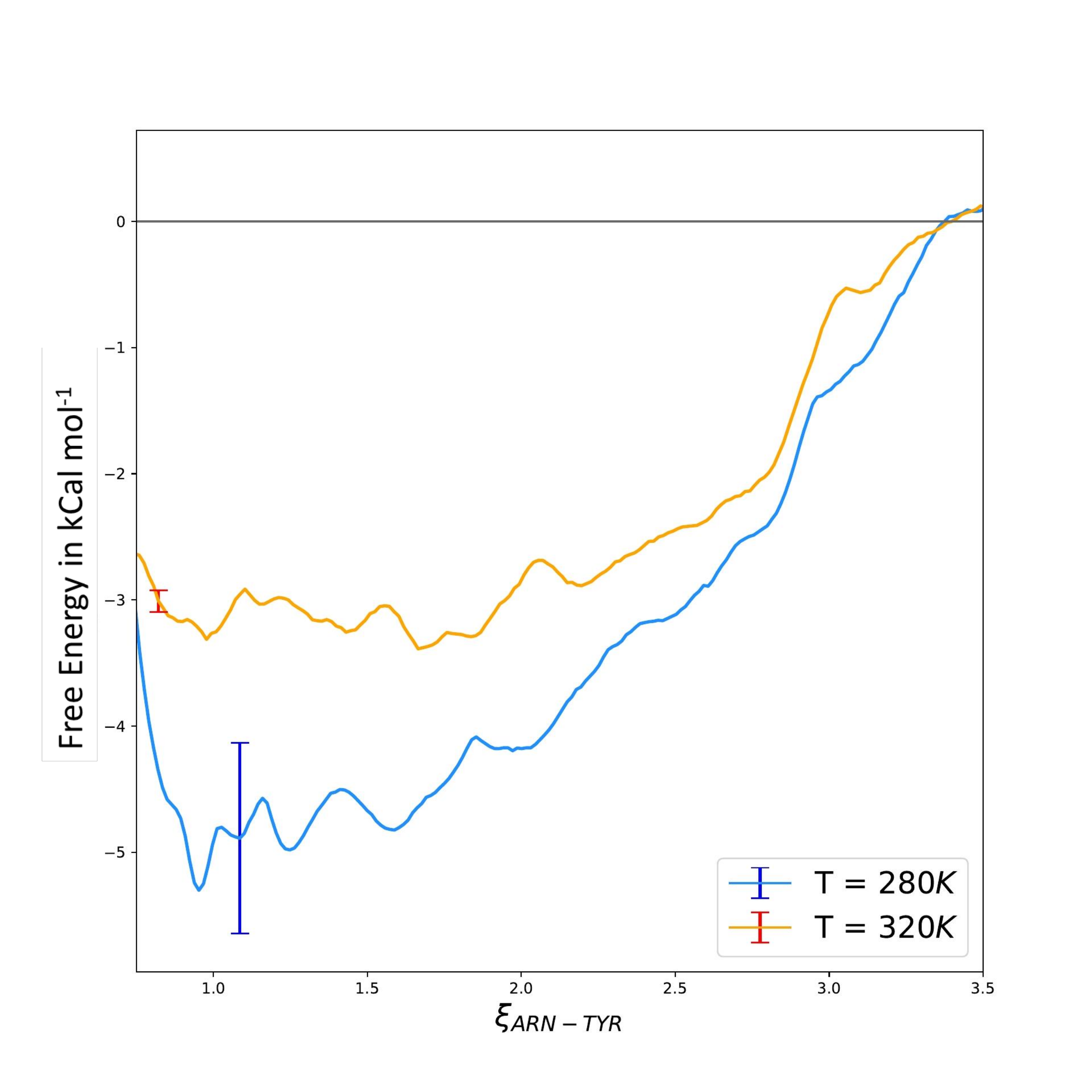}
    \hfill
    \includegraphics[width=0.475\linewidth]{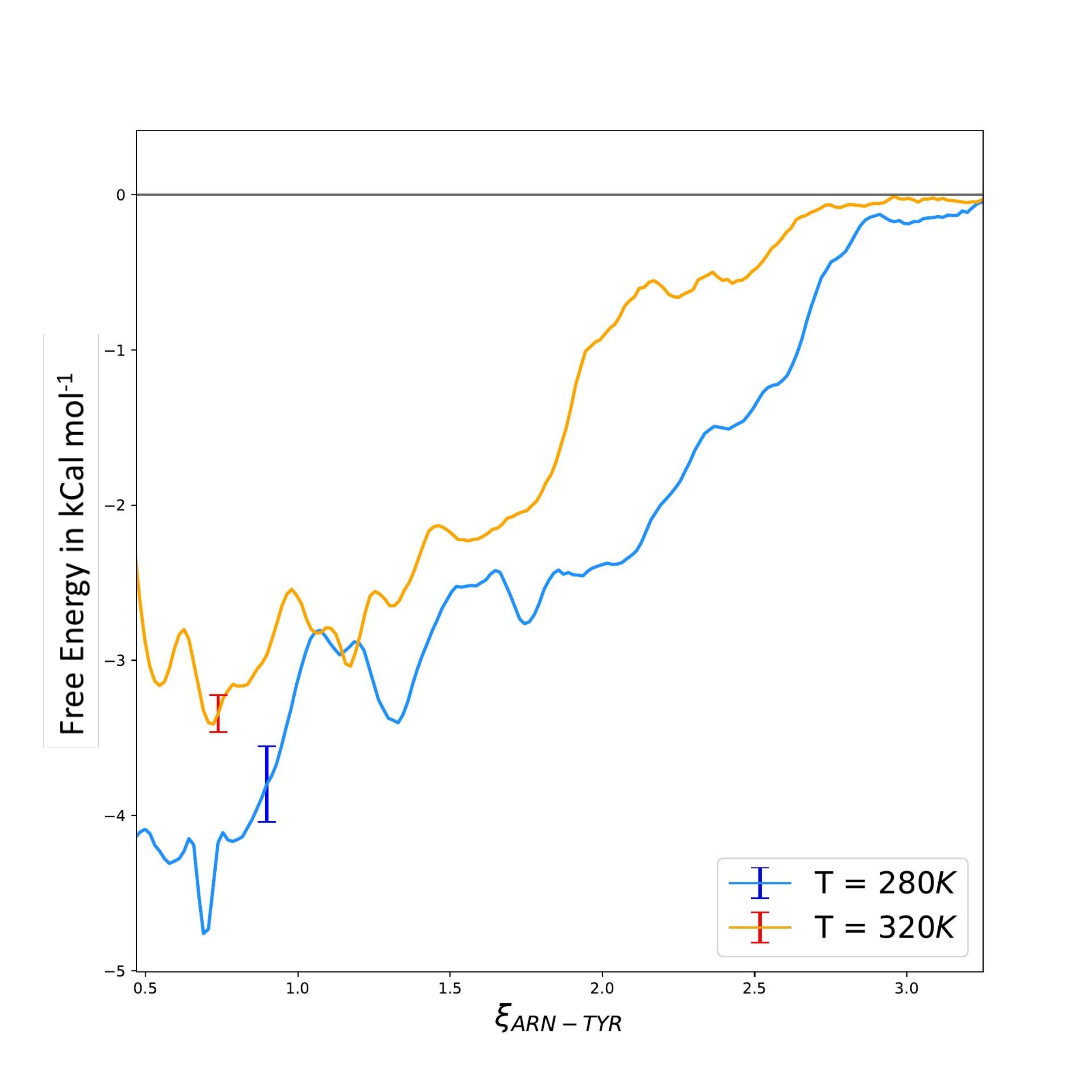}
    \caption{PMF for the complexation of poly-arginine and poly-tyrosine (\textbf{pR-pY}) in the absence of added salt (left) and in in the presence of 0.1M added salt (right).}
  \label{fig:diff}
\end{figure}

The association of a pR and a pY oligomer is favorable and energetically driven with and without added salt.  Figure~\ref{fig:diff} depicts the PMF between pY-pR oligomers with no added salt and 0.1M added salt.  In both cases, and at both temperautres the PMF is negative, and of similar magnitude to the PMF between pR and pY, i.e. $\sim$ -4 kcal/mol.
In addition, the decrease in temperature results in a decrease in the free energy (more negative) which means the complexation is energetically driven in both cases.

The strength of the free energy of complexation between pR and pY is almost an order of magnitude lower than the complexation of oppositely charged polyelectrolytes and polypeptides. For instance, comparing complexation of pR-pY to that of pK-pE, a system that is known to undergo complex coacervation, reveals that the free energy of association of pK-pE is approximately 10 kcal mol$^{-1}$ more favorable.\cite{singh2020driving}  This questions the notion that LLPS proceeds first via the complexation of R and Y rich peptide fragments via $\pi$-cation bonding between arginine and tyrosine residues.  The free energy of association is no stronger than it is for the association between two pY oligomers.

The pR-pY complexation is energetically driven in contrast to complexation of oppositely charged polyions, which is 
strongly entropy-driven and  energetically unfavorable.\cite{priftis2012thermodynamic,singh2020driving}. Furthermore, the release of counterions into solution during association, a mechanism that is known to drive polyelectrolyte complexation, is not observed during pY-pR complexation, with or without added salt.

The presence of 0.1M salt makes pY-pR complexation less favorable at the lower temperatures. This can be seen in Fig \ref{fig:diff}. The calculations are qualitatively similar to the system without salt, in that the compelexation is both favorable at both temperatures and is energetically driven. Quantitatively, there is a decrease of $\leq$ 1 kcal mol$^{-1}$ in the magnitude of free energy at T=280K when salt is added. As a result, the energetic interactions between pR and pY decrease with the increase in salt concentration.

\begin{figure}[H]
  \centering
    \includegraphics[width=0.475\linewidth]{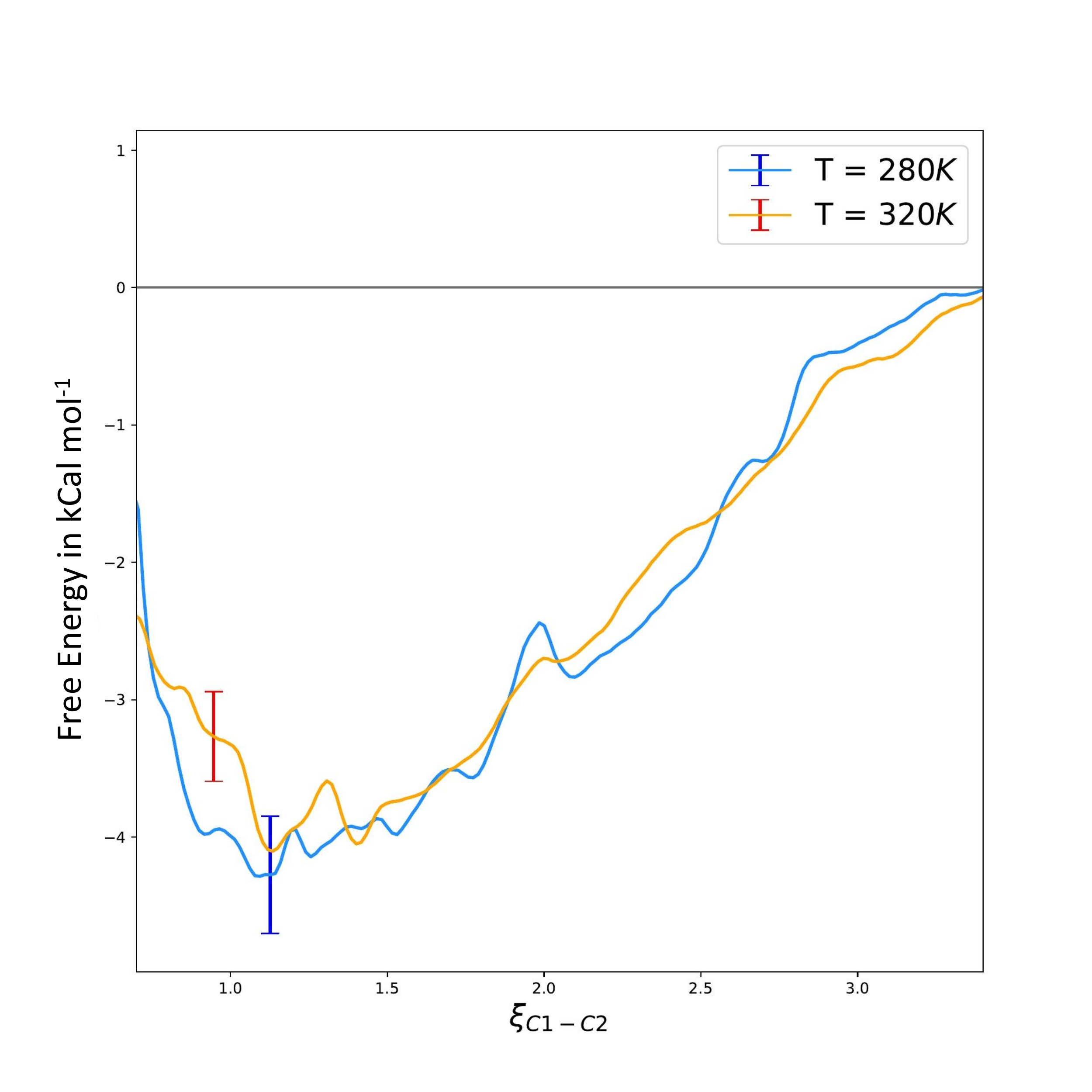}
    \caption{PMF between two neutral complexes, each made up of a pair of poly-lysine and poly-glutamate. The coordinate ($\xi_{C1-C2}$) can be thought of as one that encapsulates the second binding step of complex coacervation, where multiple poly-electrolyte pairs aggregate to form a coacervate (or a polymer rich) phase.}
  \label{fig:combined}
\end{figure}

The complexation of polyions of opposite charge is strongly entropy driven\cite{singh2020driving} in contrast to the energy driven association between pY-pY and pR-pY.  It has been suggested that the phase separation in coacervation is similar to a phase separation in neutral polymer solutions, and we therefore investigate the association of neutral (complexed) oligomers.  Each complex consists of one oligomer of poly-lysine (pK) and one oligomer of poly-glutamate (pE), as visualized in Fig \ref{fig:4_chains_visual}. The PMF is shown in \ref{fig:combined} for two temperatures.  From the temperature dependence we find that the complexation is energy driven.  The magnitude of the association and location of the minimum are both quite similar to what was seen for pY-pR and pY-pY suggesting that the LLPS could be similar to the second step of coacervation.  The magnitude of the free energy and the energetic driving force are consistent with the empirical findings of  Prifthis et al. \cite{priftis2012thermodynamic}. 

We therefore conclude that complex coacervation could occur by an entropy-driven complexation and energy-driven phase separation.  However, this mechanism cannot be invoked for systems driven by $\pi$-cation bonds between arginine and tyrosine where the minimum is shallow enough that the constituents cannot be considered complexed. Only the aggregation step of complex coacervation is applicable to systems that show $\pi$-cation mediated phase separation.

\subsection{$\pi$-Cation Interactions} \label{ssec:picat}

\begin{figure}[H]
\includegraphics[width=0.8\linewidth]{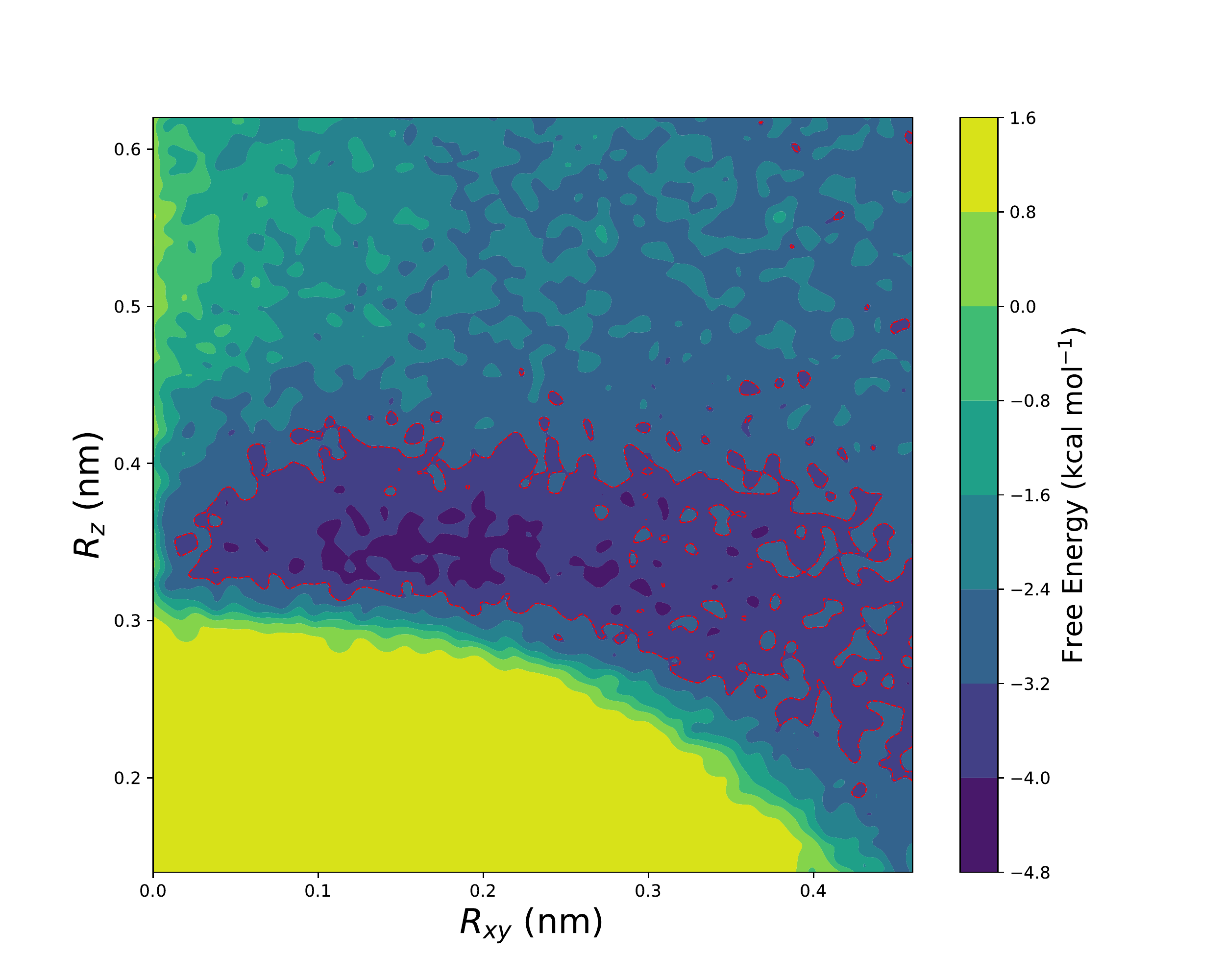}
\caption{Free Energy Surface obtained from well-tempered metadynamics along the components $R_z$ and $R_{xy}$ axis (described in figure~\ref{fig:picat_visual}), of distance between the arginine carbocation and the center of mass of the aromatic ring of tyrosine. The region within the red contour lines, which marks the regions with free energy $\leq$ -3.2 kcal mol$^{-1}$ is chosen as the quantitative measure for determining the existence of a $\pi$-cation bonds.}
  \label{fig:metadynamics}
\end{figure}

In order to determine if cation-$\pi$ interactions are important we obtain a geometric criterion for this interaction in solution.  Note that in atomistic force fields these interactions are not explicitly incorporated (similar to the case for hydrogen bonds) but are implicitly present in the fit of force field parameters to structural and thermodynamic data.  To obtain a geometric criterion for $\pi$-cation interactions we perform well tempered metadynamics for single capped R and Y amino acids in solution.  Figure~\ref{fig:metadynamics} depicts the free energy as a function of the co-ordinates $R_z$ and $R_{xy}$ (see 
figure~\ref{fig:picat_visual}).  The result obtained from the metadynamics in solution are qualitatively similar to that of Kumar et al. gas-phase quantum calculations \cite{kumar2018cation}. The only major difference in their calculation is that the interaction energy is stronger at certain regions by approximately 1-2 kcal mol$^{-1}$.  

The region within the red countour lines in Fig \ref{fig:metadynamics} is used to characterize a $\pi$-cation bond.  Space is discretized as in the metadynamics simulations, and if the $R_z$ and $R_{xy}$ between residues of arginine and tyrosine falls within the highlighted region, then it is counted as 1 $\pi$-cation bond. This analysis is done for all the pY-pR systems, however it is important to note that it is possible to have $\frac{11\times10}{2}$ $\pi$-cation bonds in total, since pR and pY have 11 residues of arginine and tyrosine respectively.

\begin{figure}[H]

  \centering
  \begin{subfigure}[b]{0.45\linewidth}
    \includegraphics[width=\linewidth]{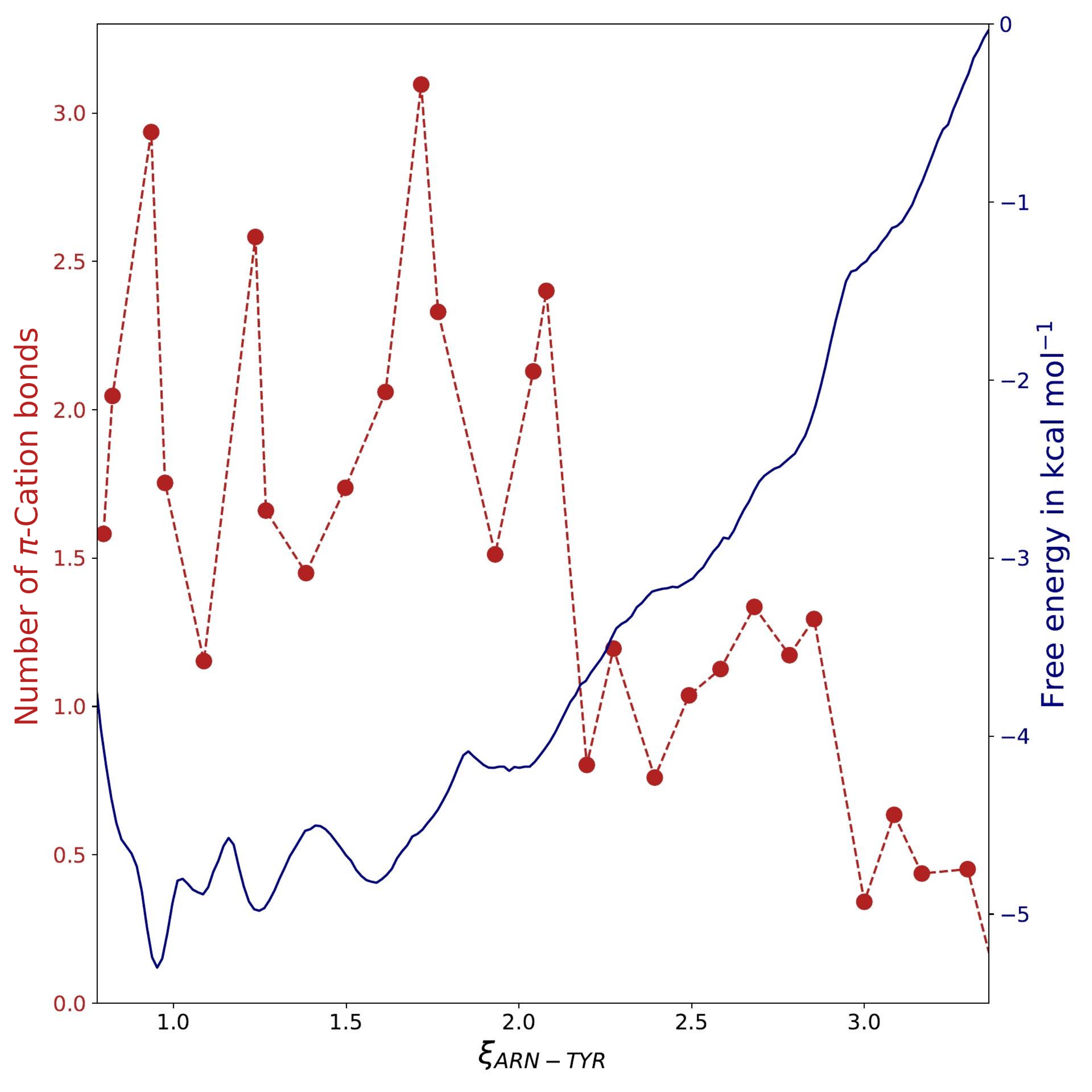}
    \caption{C$_{NaCl}^{excess}$ = 0.0M}
  \end{subfigure}
  \begin{subfigure}[b]{0.45\linewidth}
    \includegraphics[width=\linewidth]{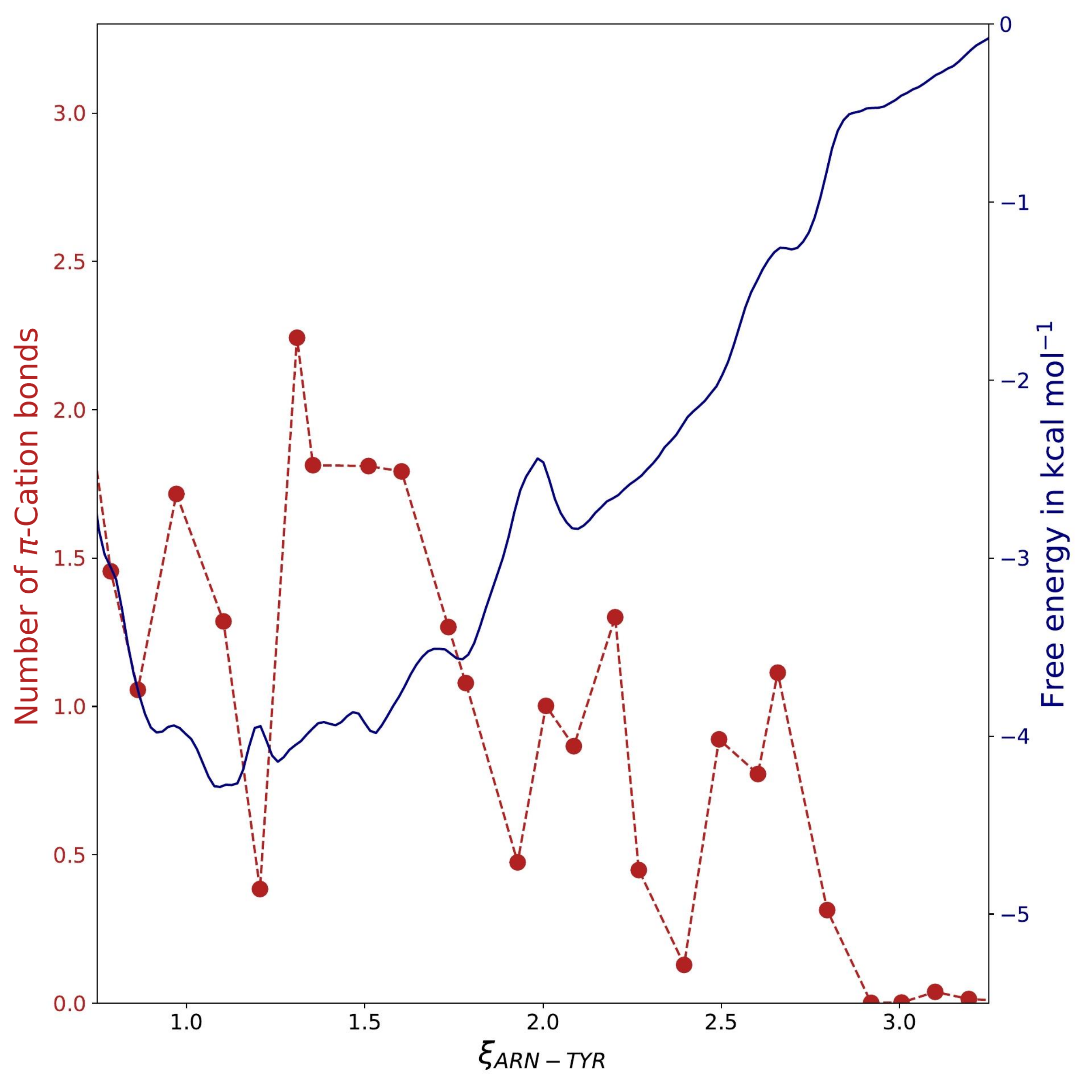}
    \caption{C$_{NaCl}^{excess}$ = 0.1M}
  \end{subfigure}
  \caption{Number of $\pi$-Cation bonds between pY and pR as a function of distance between the two oligomers ($\xi_{ARN-TYR}$) without (a) and with (b) added salt. The free energy between between the two oligomers is shown as a reference.}
  \label{fig:picat_distance}
\end{figure}

The number of $\pi$-cation bonds between pR and pY as a function of the distance between them ($\xi_{ARN-TYR}$) is shown in Fig \ref{fig:picat_distance}. The free energy is also shown in the figures as a reference. It is expected that the number of $\pi$-cation bonds should decrease and eventually go to 0 as we increase the distance between pY and pR, and this can be observed in the plots. The number of $\pi$-cation bonds goes to 0 at $\xi_{ARN_TYR}\approx 3.25$, which is also the distance at which the free energy between them plateaus.

The favorable free energy between pR and pY can almost exclusively be attributed to $\pi$-cation bonding. One reasoning for this is the fact that free energy between pR-pY plateaus when the number of $\pi$-cation bonds goes to zero. A stronger argument can be made by observing the correlation between the free energy and $\pi$-cation bonds. The free energy is not completely linear, but shows fluctuations (troughs and peaks) along the coordinate $\xi_{ARN-TYR}$. Interestingly, these fluctuations show strong correlations with the number of $\pi$-cation bonds -  distances at which the free energy is higher in magnitude compared to its surroundings also has a higher number of $\pi$-cation bonds and vice versa. This can be seen clearly by focusing on the $1.0 \leq \xi_{ARN-TYR} \leq 2.0$ region.

\begin{figure}[H]
\includegraphics[width=0.8\linewidth]{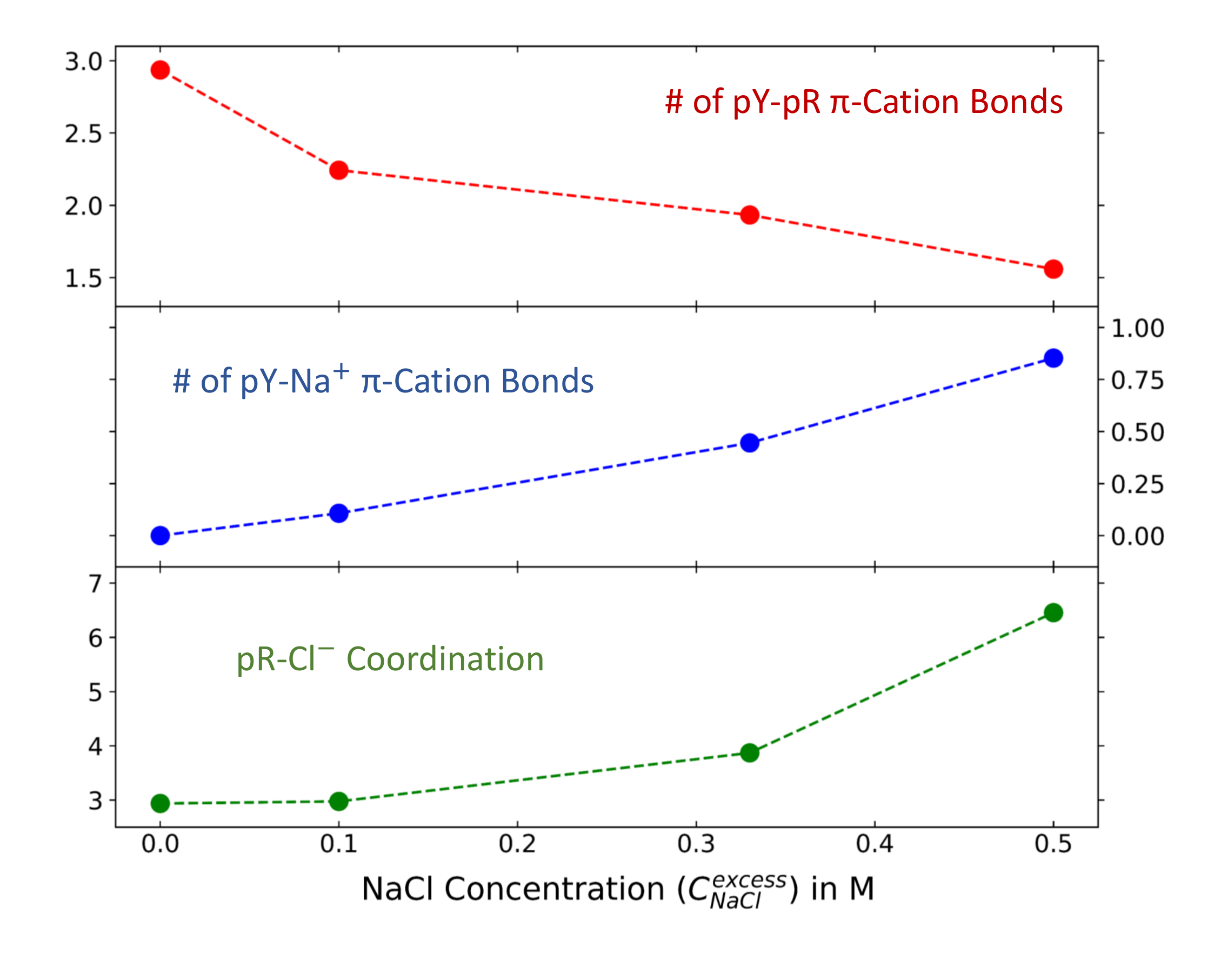}
\caption{Effect of the concentration of salt on the number of $\pi$-cation bonds between pY-pR (top), number of $\pi$-cation bonds between pY-Na$^+$ (middle) and number of Cl$^-$ ions within the second shell of the arginine carbocation.}
  \label{fig:salt}
\end{figure}

The number of $\pi$-cation bonds formed between pY and pR is  reduced with the addition of salt, by approximately 1 with the addition of 0.1M salt. This implies that the small anions (Cl$^-$) either shield the electrostatic interactions of the $\pi$-cation bonds between R and Y residues, or the Na$^+$ counterions compete with the R carbocation and form $\pi$-cation bonds with Y residues. To further explore the effect of the salt concentration on $\pi$-cation bonds between R and Y residues, 4 simulations are run with salt concentrations of 0M, 0.1M, 0.33M and 0.5M. The number of $\pi$-cation bonds between pR and pY, $\pi$-cation bonds between Na$^+$, and the number of $Cl^{-}$ ions within the second shell of the pR carbocation is plotted in Fig \ref{fig:salt}.

This plot clearly reveals that the number of pR-pY $\pi$-cation bonds show a strong negative correlation with the number of pY-Na$^+$ $\pi$-cation bonds and pR-Cl$^-$ coordination number. We deduce that there are two major mechanisms are responsible for the salting-in behavior observed in LLPS. With the increase in salt concentration, the Na$^+$ ions replace the R carbocations and form $\pi$-cation bonds with Y residues - this is evident from the increase in the average number of pY-Na$^+$ $\pi$-cation bonds as the salt concentration is increased. Secondly, the number of Cl$^-$ within the second shell of the pR carbocation increases with the salt concentration, indicating that the electrostatic screening effect due to Cl$^-$ ions destabilize the pR-pY $\pi$-cation bond at high salt concentrations.

\section{Conclusion}

We calculate the free energy of association between oligomers containing residues of arginine (pR) and tyrosine (pY). These calculatons reveal that pR-pY complexation is energy-driven.  This is fundamentally different from complexation of charged polyelectrolytes, which is much more strongly favorable and entropy-driven. 

The association of two polyelectrolyte complexes, however, displays a free energy profile that is very similar to that of pR and pY.  This suggests a striking similarity between pR-pY complexation and the second binding step of complex coacervation, in which neutral polylelectrolyte pairs aggregate together to form a dense and dilute phase. We finally conclude that complex coacervation as a whole is not invoked by systems that phase separate through $\pi$-cation bonding. Rather, the aggregation step of complex coacervation by itself is the correct mechanism for these systems.

Finally, we  find that the energetic contribution of pR-pY complexation comes primarily from $\pi$-cation bonding. On adding excess salt, these interactions are weakened by two effects - screening by the small anions and competition of the arginine carbocation with the small cations to form $\pi$-cation bonds.

The results suggest that the complexation of peptides leading to LLPS could be rather different from the coacervation oppositely charged polyions.  However, if the latter is seen as two steps, a complexation of polyions followed by a phase separation of neutral moities, then the second step in the thermodynamic cycle is similar to the association of peptides.

\section{Author Information}
\textbf{Corresponding Author}\\
*E-mail: \href{yethiraj@chem.wisc.edu}{yethiraj@wisc.edu}\\
\textbf{ORCID}\\
Aditya N. Singh: \href{https://orcid.org/0000-0002-8019-2967}{0000-0002-8019-2967}\\
Arun Yethiraj: \href{http://orcid.org/0000-0002-8579-449X}{0000-0002-8579-449X}\\
\textbf{Notes}\\
The authors declare no competing financial interest.
\begin{acknowledgement}
This research work was supported by the National Science Foundation through Grant No. CHE-1856595. The authors would like to thank Dr. Ajay Muralidharan for insightful discussion regarding the metadynamics calculation performed in this study. All simulations presented here were performed in computational resources provided by UW-Madison Department of Chemistry HPC Cluster under NSF Grant No. CHE-0840494, the UW-Madison Center for High Throughput Computing (CHTC). The authors would also like to thank the Texas Advanced Computing Center (TACC) at The University of Texas at Austin for providing HPC resources that have contributed to the research results reported within this paper (URL: http://www.tacc.utexas.edu). 
\end{acknowledgement}

\bibliography{Coacervates}

\end{document}